# THE SUN IS A PLASMA DIFFUSER
# THAT SORTS ATOMS BY MASS

©2005 O. Manuel[1], S. A. Kamat[1], M. Mozina[2]

[1]*University of Missouri, Rolla, USA* omatumr@yahoo.com
[2]*Emerging Technologies, Mt. Shasta, CA, USA* michael@etwebsite.com

The Sun is a plasma diffuser that selectively moves light elements like H and He and the lighter isotopes of each element to its surface. The Sun formed on the collapsed core of a supernova (SN) and is composed mostly of elements made near the SN core (Fe, O, Ni, Si, and S), like the rocky planets and ordinary meteorites. Neutron-emission from the central neutron star triggers a series of reactions that generate solar luminosity, solar neutrinos, solar mass-fractionation, and an outpouring of hydrogen in the solar wind. Mass fractionation seems to have operated in the parent star, and likely occurs in other stars as well.

## INTRODUCTION

In 1913 Aston [1] produced neon of light atomic weight by diffusion. He later used electric and magnetic fields to measure abundances and masses of isotopes [2]. In 1969 lightweight neon from the Sun was discovered in lunar soils [3]. Recent measurements with modern mass spectrometers and "running difference" images of the Sun have uncovered this surprising record of the Sun's origin, composition, and operation:

- *Isotope analyses* revealed extinct nuclide decay products [4, 5], isotope variations from nucleosynthesis [6-9], and multi-stage fractionation in the Sun [10, 11].

- *Nuclide masses* showed repulsive n-n interactions, high potential energy for those in a neutron star, and a source for luminosity, neutrinos, and the carrier gas that sustains mass fractionation and an outflow of solar-wind hydrogen [12, 13].

- *"Running difference" images* of the Sun with filters to enhance emissions from Fe (IX) and Fe (X) show a rigid, Fe-rich surface [14] beneath the Sun's fluid photosphere of lightweight elements [15].

Readers may wish to review Figs. 1-8 and seek other explanations if confused by this story connecting decades of complex data to a few simple conclusions.

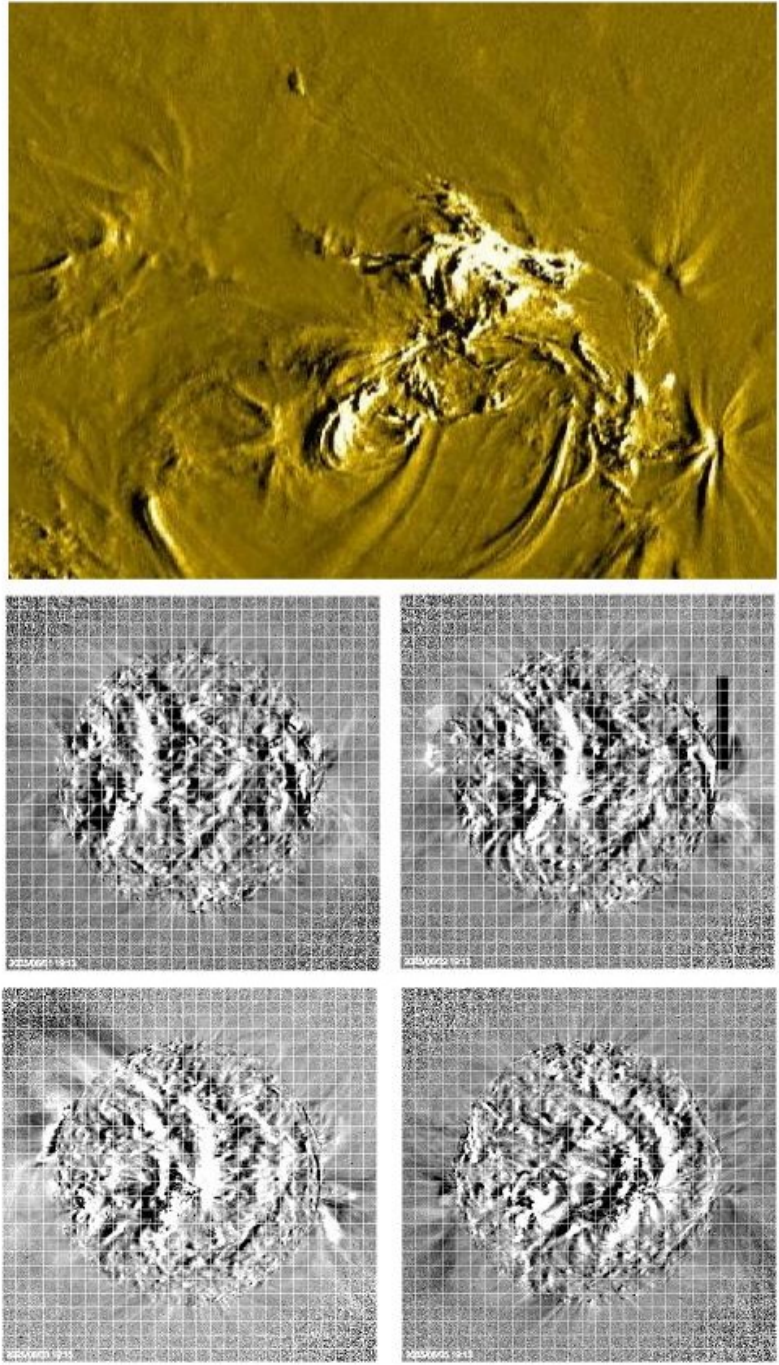

Fig 1. <u>The top section</u> is a "running difference" image of the Sun's iron-rich sub-surface from the Trace satellite using a 171 Å filter sensitive to Fe (IX) and Fe (X) emissions. A movie of a flare and mass ejection from this region of AR 9143 on 28 August 2000 is here: http://vestige.lmsal.com/TRACE/Public/Gallery/Images/movies/T171_000828.avi <u>The bottom section</u> has a grid system to show rotation (left to right) of the Sun's rigid, iron-rich structure over a 5-day period of 1-5 June 2005 in four images from the SOHO satellite using a 195 Å filter that is also sensitive to Fe (XII) emissions.

New satellite systems specifically designed to study solar activity allow observations of the transitional region of the Sun with remarkable precision and provide unprecedented detail on solar activity. Images from the SOHO and Trace satellite programs reveal rigid, persistent structures in the transitional layer of the Sun, in addition to its dynamic, fluid photosphere. While looking at those images, Mozina [14] made a startling discovery: *"After downloading a number of these larger "DIT" (grey) files, including several "running difference" images, it became quite apparent that many of the finer details revealed in the raw EIT images are simply lost during the computer enhancement process that is used to create the more familiar EIT colorized images that are displayed on SOHO's website. That evening in April of 2005, all my beliefs about the sun changed."*

Fig. 1 shows the images he observed. The top section is a "running difference" image of the Sun's iron-rich, sub-surface revealed by the Trace satellite using a 171 Å filter. This filter is sensitive to emissions from Fe (IX) and Fe (X). Lockheed Martin made this movie of the C3.3 flare and a mass ejection in AR 9143 from this region on 28 August 2000. http://vestige.lmsal.com/TRACE/Public/Gallery/Images/movies/T171_000828.avi

The bottom of Fig. 1 shows four images taken over a 5-day period on 1-5 June 2005 of a rigid, iron-rich structure below the Sun's fluid photosphere. These "running difference" images from SOHO used a 195 Å filter to enhance light emissions from Fe (IX) and Fe (X). Videos of these images show the rotation (from left to right for the images in the bottom part of Fig. 1) that led Mozina to conclude that the Sun's iron-rich sub-surface rotates uniformly, from pole to equator, every 27.3 days [14].

Further discussion of these images will be postponed until the experimental basis has been presented for concluding that the Sun acts as a plasma diffuser, hiding its iron-rich interior beneath a veneer of lightweight elements.

Decay products of these short-lived nuclides in meteorites (in order of decreasing half-lives) provided the first evidence that highly radioactive material formed the solar system: $^{244}$Pu ($t_{1/2}$ = 80 Myr) [5], $^{129}$I ($t_{1/2}$ = 16 Myr) [4], $^{182}$Hf ($t_{1/2}$ = 9 Myr) [16], $^{107}$Pd ($t_{1/2}$ = 6.5 Myr) [17], $^{53}$Mn ($t_{1/2}$ = 3.7 Myr) [18], $^{60}$Fe ($t_{1/2}$ = 1.5 Myr) [19], $^{26}$Al ($t_{1/2}$ = 0.7 Myr) [20], and $^{41}$Ca ($t_{1/2}$ = 0.1 Myr) [21]. A supernova likely produced these nuclides. Two of them, $^{244}$Pu and $^{60}$Fe, could only have been made in a supernova [22]. Decay products of extinct $^{244}$Pu and $^{129}$I have also been identified in the Earth [23].

By 1961 Fowler *et al.* [24] noted that the levels of short-lived radioactivity were higher than expected if an interstellar cloud formed the solar system. The discrepancy between isotope measurements and the nebular model for formation of the solar system increased dramatically after nucleogenetic isotopic anomalies [6-9] and the decay products of even shorter-lived nuclides were discovered in meteorites [16-21].

Combined Pu/Xe and U/Pb age dating showed that the $^{244}$Pu was produced ≈5 Gyr ago in a supernova explosion [25]. Age dating with $^{26}$Al/$^{26}$Mg showed that some refractory meteorite grains started to form within ≈ 1-2 Myr after the explosion [26].

All primordial He was linked to excess $^{124}$Xe and $^{136}$Xe when meteorites formed [7-8, 27-28]. That was the first indication that fresh supernova (SN) debris directly formed the solar system (Fig 2), before elements and isotopes in different SN layers mixed.

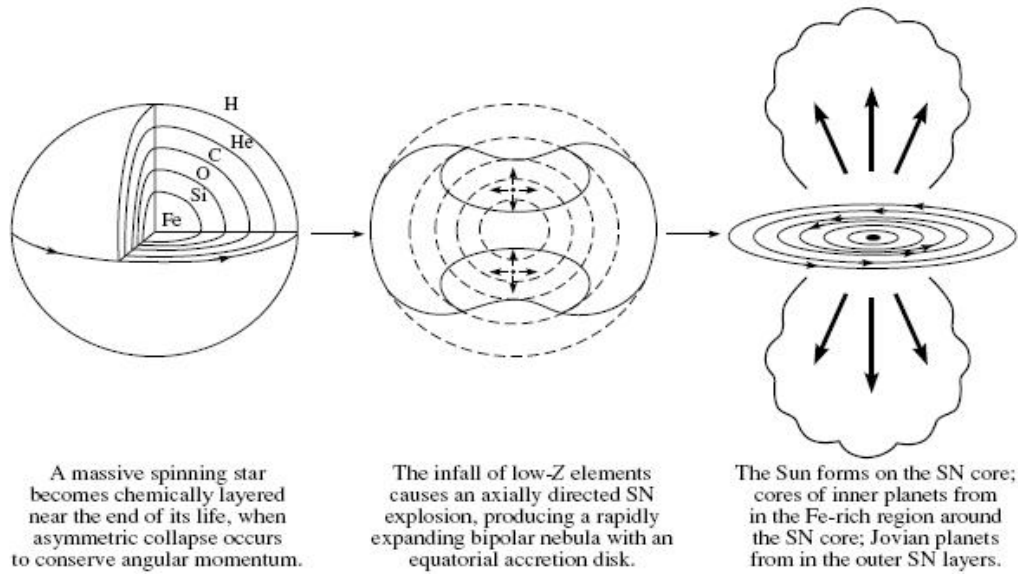

A massive spinning star becomes chemically layered near the end of its life, when asymmetric collapse occurs to conserve angular momentum.

The infall of low-Z elements causes an axially directed SN explosion, producing a rapidly expanding bipolar nebula with an equatorial accretion disk.

The Sun forms on the SN core; cores of inner planets from in the Fe-rich region around the SN core; Jovian planets from in the outer SN layers.

Fig 2. Fresh, radioactive debris of a supernova that exploded here ≈ 5 Gyr ago [25] started to form refractory grains in the solar system within ≈ 0.001-0.002 Gyr [26], before elements and isotopes from the different SN layers had completely mixed [7, 8, 27, 28].

Excess r- and/or p-products [22] were found in the isotopes of Xe [6], Kr [29], and Te [8] in some meteorite minerals. The middle isotopes of Xe [30], Kr [30] and Te [8] were found to be enriched in other meteorite minerals from the s-process of nucleosynthesis [22].

By 1993 analyses had also revealed excess r- and p- products in the isotopes of Ba, Nd, and Sm in some meteorite minerals, and excess s-products in others [31]. Since excess r- and p-products can otherwise be considered as a deficit of s-products, and excess s-products can instead be viewed as deficits of p- and r-products, Begemann [31] noted that these "mirror-image" (+) and (-) isotopic anomalies may be separate products from the stellar nuclear reactions that collectively produced "normal" isotope abundances.

Fig. 2 was posited [7, 8] to explain the link of primordial He and Ne with "strange" Xe ("strange" isotope abundances) in meteorites, and their absence in the noble gas component with "normal" Xe isotope abundances [27, 29, 32].

Nuclear reactions made different isotopes, at different times, in different stellar layers [22]. This record was not destroyed in the SN explosion i.e., *the neutron flux that made r-products did not permeate the entire star*. Isotope and element variations were linked in the parent star [22], and they remained linked together over planetary distances as the heterogeneous SN debris formed the solar system. Thus, major elements formed host minerals whose average atomic number ($\bar{Z}$) increased with stellar depth, trapping the isotopic anomalies generated in that region.

a.) (C, $\bar{Z} = 6$): In the outer SN layers the r- and p-processes made the "strange" Xe that became trapped with primordial He and Ne [7] in diamond inclusions of meteorites [29, 33-35]. Heavy and light isotopes of other elements are enriched in the diamonds [8, 36, 37]. The Galileo probe also found this same "strange" Xe in Jupiter's He-rich atmosphere [38], as expected from the scenario shown in Fig 2.

b.) (SiC, $\bar{Z} = 10$): Deeper in the SN, in a region less altered by the r- and p-processes [22], SiC trapped excess $^{22}$Ne from mass fractionation [39] with excess middle isotopes (s-products) of Xe, Kr, Te, Ba, Nd and Sm [8, 30, 31, 40, 41]. SiC is also the likely carrier of s-products just found in Os from unequilibrated chondrites [42].

c.) ($SiO_2$, $\bar{Z} = 11$): Silicates are abundant and show few anomalies in meteorites. A component of *"almost pure $^{16}O$"* [ref. 43, p. 485] was reported in carbonaceous stone meteorites in 1973. In 1976 it was noted that the six classes of meteorites and planets each have characteristic levels of excess $^{16}O$ [44]. "Strange" isotope ratios in a silicate particle of interplanetary dust were recently cited as evidence of a probable supernova origin [45]. Isotopic anomalies and the decay product of $^{27}$Al were found in spinel (MgAl$_2$O$_4$, $\bar{Z}$ = 11) meteorite grains. These findings require RGB or AGB stars with *"hot bottom burning"* and *"cool-bottom processing"* [46] if the spinel grains are not products of local element synthesis, as shown in Fig. 2.

d.) (Fe, Z = 26): Deep in the SN debris, iron meteorites and the cores of the terrestrial planets formed. Iron meteorites trapped "normal" Xe, like that on Earth [47]. The University of Tokyo [48], Harvard [49] and Cal Tech [50] have new data showing that iron meteorites did not form by the extraction of iron from an interstellar cloud. *The stable isotopes of molybdenum made by different stellar nuclear reactions (e.g., $^{92}$Mo from the p-process, $^{96}$Mo from the s-process, $^{100}$Mo from the r-process) are not completely mixed, even in massive iron meteorites* [48-50]!

e.) (FeS, $\bar{Z}$ = 21): Between silicates and iron, "normal" Xe and other elements were trapped in troilite of meteorites and in Fe,S-rich planets like Earth and Mars [51, 52]. "Normal" xenon is also in the solar wind but the lighter mass Xe isotopes are enriched by about 3.5% per mass unit [23, 53], as will be discussed next.

UBIQUITOUS MASS FRACTIONATED ISOTOPES

In 1960 Reynolds noted that mass fractionation might explain differences between Xe isotopic compositions in meteorites and in air [54]: *"The xenon in meteorites may have been augmented by nuclear processes between the time it was separated from the xenon now on earth and the time the meteorites were formed"*, or *"On the other hand a strong mass-dependent fractionation may be responsible for most of the anomalies"* [p. 354].

The Xe isotope data [54] required many fractionation stages (≈ 10). Fractionation was later seen in He, Ne, Ar, Kr and Xe isotopes, but *the fractionation site was not identified* [55-63]. Lighter elements showed larger variations, as expected, but doubts about fractionation prevailed [64-67] and variations in the isotope abundances of He and Ne were instead labeled alphabetically as distinct primordial components [64-74].

Fig. 3 shows that the Ne isotopes in air, in the solar wind (SW) and in the gas released from the Fayetteville meteorite [56] can be explained as mixtures of cosmogenic neon and mass fractionated neon lying along the dashed line. Neon isotopes in carbonaceous chondrite meteorites (not shown) also lie on the fractionation line at $^{20}$Ne/ $^{22}$Ne = 8, but *a possible site for ≈ 10-stages of diffusive mass fractionation was unknown in 1967* [56].

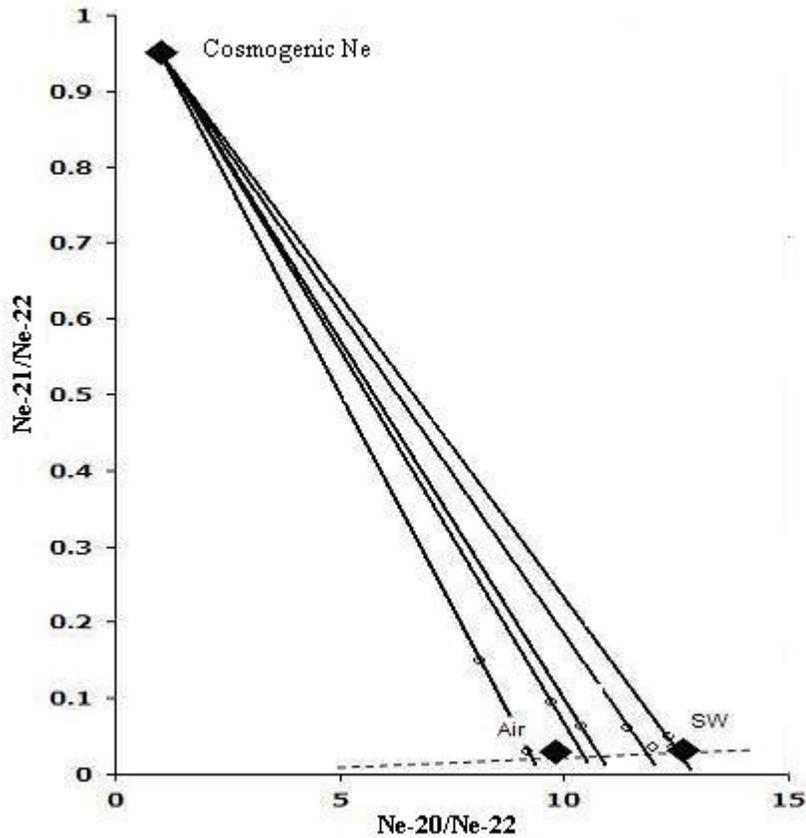

Fig 3. Neon in air, in the solar wind (SW), and that released by stepwise heating of the Fayetteville meteorite (small open circles) can be understood as a mix of cosmogenic neon (top, left) from high-energy, cosmic-ray-induced spallation reactions with mass-fractionated primordial neon lying along the dashed line [56]. Large filled diamonds identify **Cosmogenic**, **Air**, and **SW** (solar-wind) neon. Bulk neon in carbonaceous chondrites (not shown) lie on the fractionation line at $^{20}$Ne/ $^{22}$Ne = 8.

Fractionation produced smaller effects in heavy elements, like krypton and xenon [58, 61, 62], as shown in Fig. 4 for the six Kr isotopes and the nine Xe isotopes released from lunar soil sample #15601.64 [62]. Isotopes of the same mass number, A = *i*, in air (AIR) and in average carbonaceous chondrite (AVCC) meteorites are shown for comparison.

Most solar wind Kr and Xe isotopes in lunar soil sample #15601.64 lie along the solid fractionation lines passing through AIR [62]*, but the fractionation site was still unknown*. Kr and Xe in carbonaceous chondrites (AVCC), in air and in the solar wind also lie along these lines, except at $i$ = 129, 134 and 136, where radiogenic $^{129}$Xe in air [23] and excess $^{136}$Xe and $^{136}$Xe from the r-process [6] shift the data away from the fractionation lines.

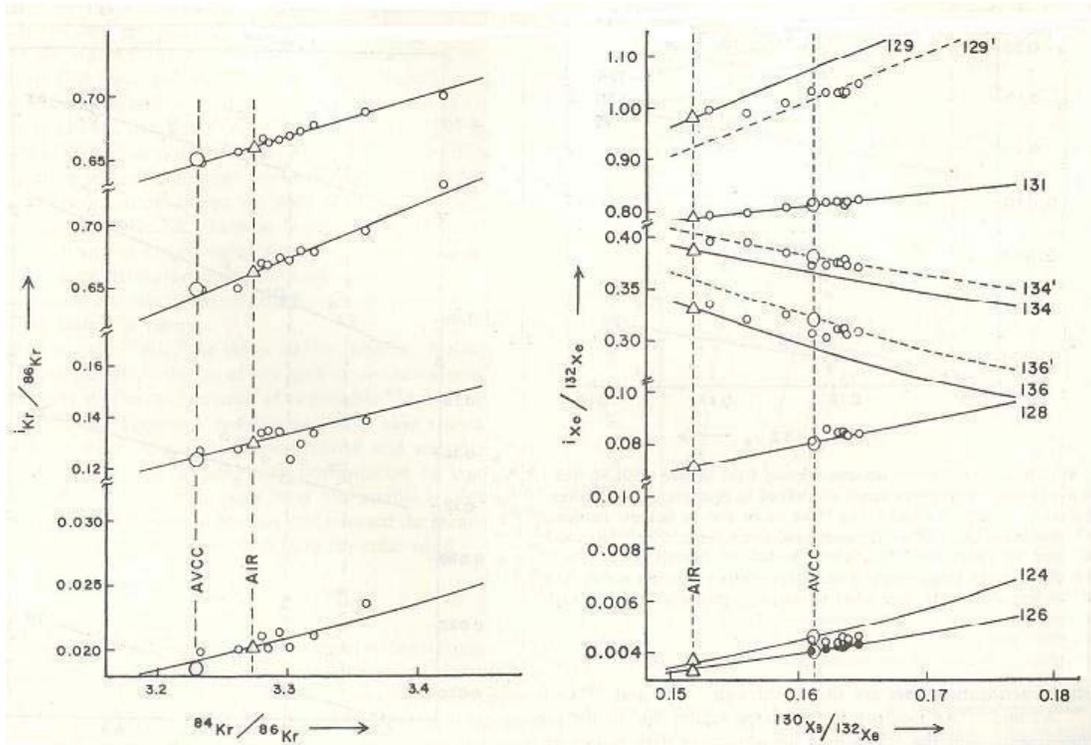

Fig 4. Solar-wind-implanted Kr and Xe isotopes in lunar soil sample #15601.64 lie along the solid mass fractionation lines that pass through AIR, except at $i$ = 129, 134 and 136, where radiogenic $^{129}$Xe in air [23] and excess $^{136}$Xe and $^{136}$Xe from the r-process [6] shift the data away from the fractionation lines [62]. Large open triangles (Δ) and large open circles (O) represent Kr and Xe isotopes in AIR and in average carbonaceous chondrite (AVCC) meteorites. Small circles (o) show solar-wind-implanted Kr and Xe isotopes in lunar soil #15601.64, corrected for products of cosmic-ray spallation reactions [62].

Large Ne isotope variations in meteorite and lunar samples [64-74] were attributed to primitive components and labeled alphabetically. Neon trapped with s-products in SiC [30], Ne-E, was reported to be almost pure $^{22}$Ne, the heaviest neon isotope [73]. However a 1980 review of Ne isotope data found that *simple mixtures of mass-fractionated and cosmogenic neon could explain all "primitive" neon components in meteorites and differences between the isotopic compositions of bulk neon in air, in the solar wind and in meteorites.* The site of such severe mass fractionation remained elusive in 1980 [75].

Upper limits on $^{20}Ne/^{22}Ne$ and $^{21}Ne/^{22}Ne$ ratios in Ne-E (Fig. 5) varied in exactly the manner expected from mass-dependent fractionation [59, 63, 75],

$$d \ln(^{21}Ne/^{22}Ne)/ d \ln(^{20}Ne/^{22}Ne) = 0.50 \qquad (1)$$

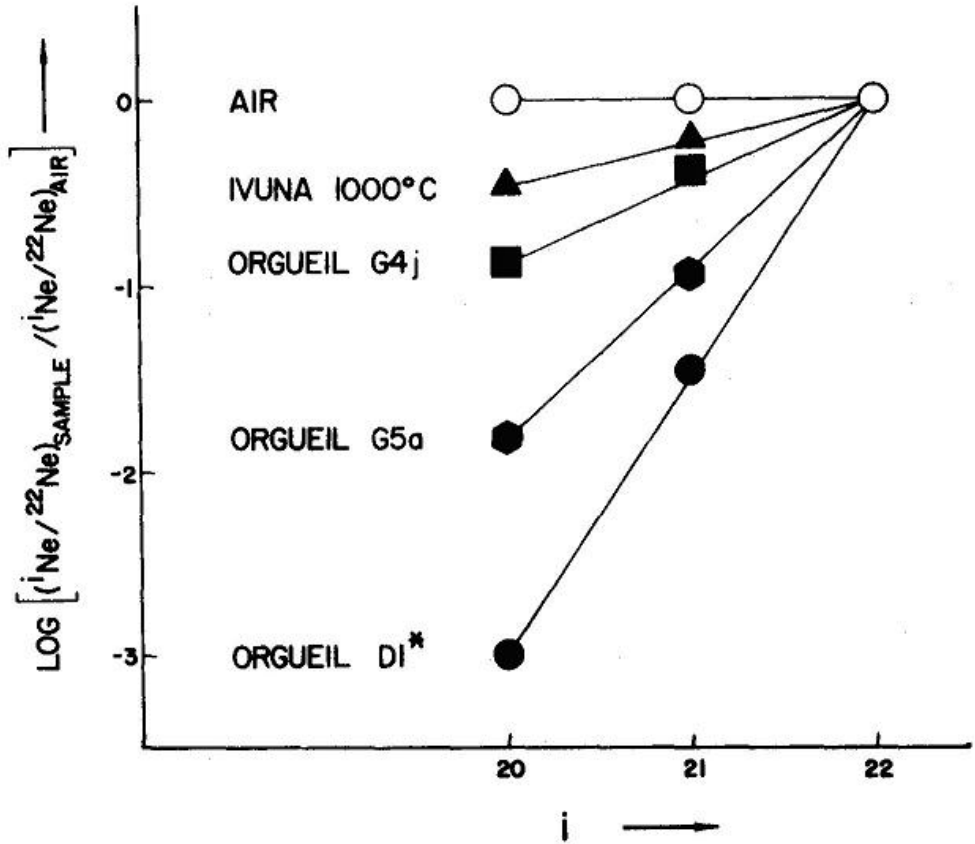

Fig. 5. Values reported as upper limits on neon isotope ratios for Ne-E in the gas released from the Ivuna carbonaceous meteorite at 1000ºC [65, 67], and in mineral separates of the Orgueil carbonaceous meteorite, G4j [69], G5a [70, 71] and D1* [72, 73] match the pattern predicted by equation (1) for multi-stage, mass-dependent fractionation.

By 1980 Clayton and Mayeda [76] had also noted *" . . . large mass fractionation of oxygen isotopes subsequent to incorporation of the nucleogenetic $^{16}O$-anomaly . . ."* and *" . . . the oxygen isotope fractionation is in constant ratio to the magnesium isotope fractionations . . ."* [76, p. 295], and Wasserburg *et al.* [77] agreed that observations on the Allende meteorite were due to *" . . . a homogenized mixture of components of extraordinary isotopic composition mixed with a component of ordinary solar system material and subjected to isotopic fractionation"*, and *"The processes responsible for* *the O and Mg nuclear effects and the astrophysical site . . . remain undefined"* [77, p. 299].

Wasserburg *et al.* [77] coined the phrase "**FUN**" to describe the **F**ractionation plus **U**nidentified **N**uclear effects in isotopes trapped in meteorites at the birth of the solar system. Fowler [78], Cameron [79] and Wasserburg [80] soon agreed that most isotope anomalies might be explained by injections of "alien" material amounting to a tiny fraction ($\approx 10^{-5}$-$10^{-4}$) of that in the solar system. However, the injection of a small amount of alien material did not explain the link between abundances of major elements with isotope anomalies, e.g., the link of primordial He with excess $^{136}$Xe from the r-process in meteorites [7, 8, 27, 32] and recently seen in Jupiter's He-rich atmosphere [38].

Manuel and Hwaung [10] took a different approach. Two types of primordial noble gases had been identified in meteorites [27]: One from the deep interior of a star contains only "normal" **Ar-1**, **Kr-1** and **Xe-1**, with isotope abundances like those on Earth and no He or Ne. The other from the outer stellar layers contains "strange" **Ar-2**, **Kr-2** and **Xe-2** and "normal" **He** and **Ne**. Manuel and Hwaung [10] *assumed that the Sun itself is a mix of these two primordial components and used isotope abundances in the solar wind to estimate the fraction of each primitive component in the Sun.*

Their comparison [10] of He and Ne isotope abundances in meteorites [27] with those in the solar wind revealed a ≈ *9-stage mass fractionation process in the Sun!* The light isotopes of He and Ne are enriched in the solar wind by nine theoretical stages of mass fractionation, $f = (H/L)^{4.5}$, each enriching the number of light mass (*L*) neon isotopes relative to that of the heavy mass (*H*) ones in the solar wind by the square root of (*H/L*).

Further, this same 9-stage fractionation process extends to the heaviest noble gas, Xe, but solar-wind xenon is mostly a mass-fractionated form of "normal" xenon (**Xe-1**), like that in air, with only a small component of "strange" xenon (**Xe-2**). Assuming that the same process sorts the intermediate noble gas isotopes, Manuel and Hwaung [10] showed that that solar krypton is a mix of the "normal" (**Kr-1**) and "strange" (**Kr-2**) seen in meteorites [27] but solar argon is the "strange" **Ar-2** that accompanies primordial He and Ne in meteorites [27]. Their results are shown on the left side of Fig. 6.

Noble gas isotopes in the solar wind reveal mass fractionation from the lightest He isotope to the heaviest Xe isotope, from A = 3 to 136 mu [10, 11]. The abundance of s-products in the photosphere offers an independent check of solar mass fractionation. The steady-flow abundance, N, of successive nuclides made by slow neutron capture [22], is inversely proportional to their neutron-capture cross-sections, σ:

$$N_{(A-1)}\sigma_{(A-1)} = N_{(A)}\sigma_{(A)} = N_{(A+1)}\sigma_{(A+1)} \tag{2}$$

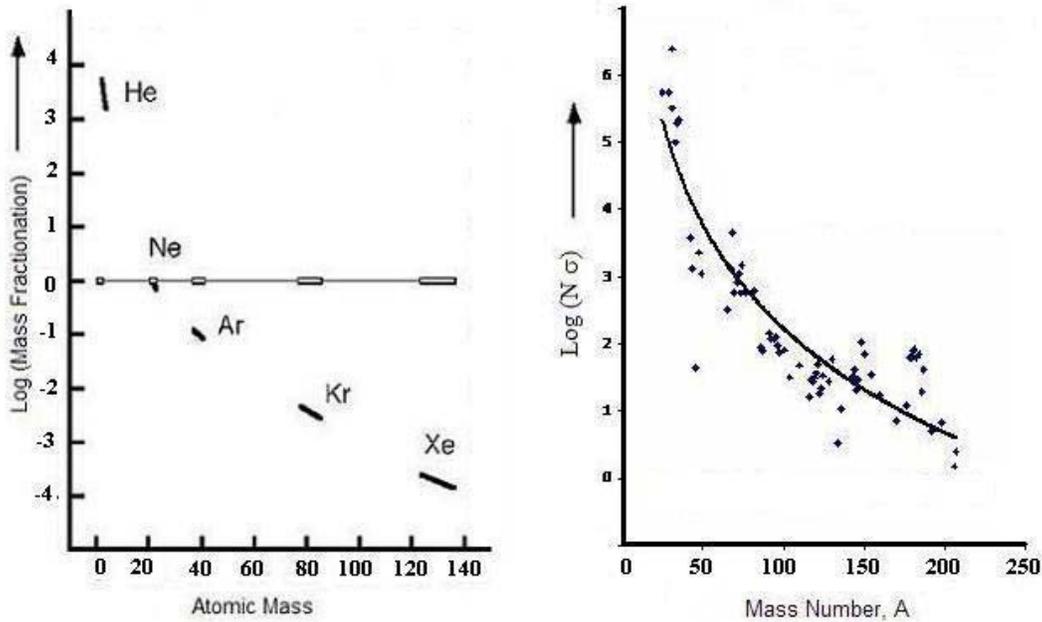

Fig. 6. <u>On the left</u>, noble gas isotopes in the solar wind (filled bars) are mass fractionated relative to those in planetary noble gases (open bars). This mass fractionation is recorded in 22 isotopes spanning a mass range of 3-136 mu [10, 11]. <u>On the right</u>, s-products in the photosphere are mass fractionated relative to the constant Nσ values expected from steady-flow [22]. This mass fractionation is recorded in the abundances of 72 s-products in the photosphere spanning a mass range of 25-207 mu [22].

Eq. (2) and steady-flow s-process have been confirmed in samarium isotopes, $^{148}$Sm and $^{150}$Sm [81], and in tellurium isotopes, $^{122}$Te, $^{123}$Te and $^{124}$Te [82], but photospheric abundances of s-products [22] exponentially decline by ~5 orders of magnitude over the mass range of A = 25-207 mu [83], as shown on the right side of Fig. 6 [83]. This confirms that fractionation occurs in the Sun itself, rather than in the solar wind.

When element abundances in the photosphere [15] are corrected for 9-stages of mass fractionation shown across the isotopes of the 22 noble gas isotopes in the solar wind (<u>left side</u>, Fig. 6), or 10 stages of mass fractionation shown across the 72 s-products in the photosphere (<u>right side</u>, Fig 6), the most abundant elements in the bulk Sun are the same: Fe, O, Ni, Si, and S. These elements all have even atomic numbers, high nuclear stability, and they are the most abundant elements in ordinary meteorites [84]. The probability of this agreement being a coincidence is essentially zero [11].

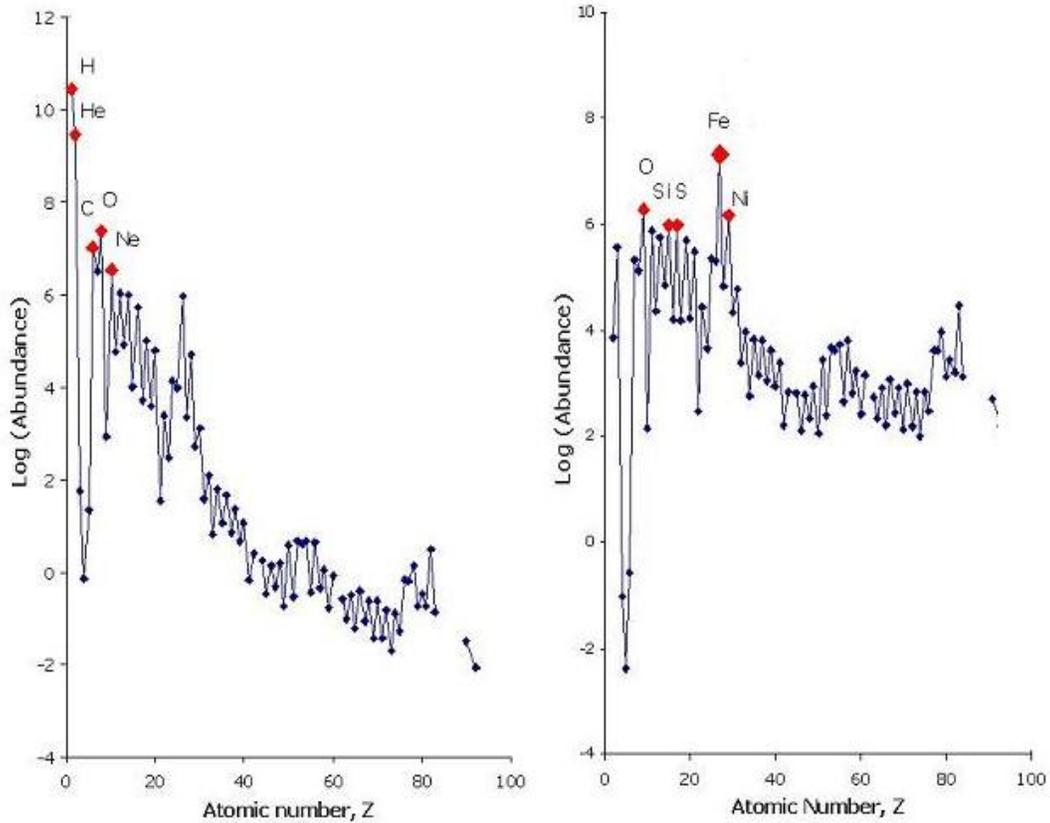

Fig. 7. <u>On the left</u>, the abundance pattern of elements reported [15] for the Sun and the solar system. <u>On the right</u>, the element abundance pattern for the Sun after correcting for the mass fractionation recorded in the isotopes of noble gases in the solar wind [10, 11].

The <u>left side</u> of Fig. 7 shows the familiar abundance pattern of elements in the solar photosphere [15]. Lightweight elements represented by large diamonds are dominant there. The <u>right side</u> of Fig. 7 shows the same abundance pattern after correcting for the mass-fractionation recorded across noble gas isotopes in the solar wind [10, 11]. These same elements are most abundant in the Sun, in rocky planets, and in meteorites.

Thus the Sun and other stars are probable sites for the mass-dependent fractionation that was repeatedly observed in isotope studies since 1960 [3, 4, 54-75]. The occurrence of this process in the parent star (See Fig. 2), as well as in the Sun, may explain why nucleogenetic isotope anomalies are embedded in elements whose isotopes have been sorted by mass, i.e., **FUN** anomalies [76, 77]. If carbonaceous chondrites formed mostly from material in the outer regions of the supernova (See Fig 2), this might also explain the similarity in the elemental abundance patterns of carbonaceous chondrites and the solar photosphere [15].

The solar abundance pattern of elements (See right side of Fig. 7) offers a viable explanation for the Sun's iron-rich, rigid structures that were shown in Fig. 1. However, iron (Fe) has tightly bound nucleons (85). The dominance of this "ash" in the Sun from fission or fusion reactions leaves the source of current solar luminosity, solar neutrinos, solar mass-fractionation, and solar wind hydrogen unexplained.

SOURCE OF SOLAR LUMINOSITY, NEUTRINOS, AND HYDROGEN

The Sun formed on the collapsed core of a supernova (Fig. 2) and consists mostly of elements (See right side, Fig**.** 7) produced in the SN interior – Fe, O, Ni, Si, and S [22]. This may seem extreme, but Hoyle [86] describes a meeting with Eddington in the spring of 1940, noting that at that time, *"We both believed that the Sun was made mostly of iron, two parts iron to one part hydrogen, more or less"*, and he continues on the same page *"The high-iron solution continued to reign supreme (at any rate in the astronomical circles to which I was privy) until after the Second World War, . . ."* [ref. 86, p. 153].

To see if some overlooked form of nuclear energy might be the source of solar luminosity (since nucleons are tightly packed in Fe, O, Ni, Si, and S [85]), students in an advanced nuclear chemistry course in the spring of 2000 were assigned the task of re-examining systematic properties of all 2,850 known nuclides [85] and using reduced nuclear variables, like the reduced physical variables that had been used in developing the corresponding states of gases [87].

After combining the nuclear charge, Z, and the mass number, A, into one reduced variable, $Z/A$ = the charge per nucleon, the values of $Z/A$ for all known nuclides lie within in the range of $0 \geq Z/A \leq 1$. After combining the atomic mass and the mass number into the reduced variable used by Aston [2], M/A = potential energy per nucleon, the values of M/A for all known nuclides lie close to the value of 1.00 mass units per nucleon. Aston [2] subtracted 1.00 from the value of each to obtain the quantity called the *"packing fraction"* or *"nuclear packing fraction."*

The <u>left side</u> of Fig. 8 shows the "cradle of the nuclides" [12, 28] that emerged when values of M/A for all 2,850 nuclides [85] were plotted against values of Z/A and then sorted by mass number, A. The <u>right side</u> of Fig. 8 shows the intercepts that mass parabolas, fitted to the data [85] at each value of A, make with the front and back planes at Z/A = 0 and Z/A = 1. At each mass number, cross-sectional cuts through the "cradle" yield values of M/A at Z/A = 0 that exceed the M/A value of a free neutron, typically by ≈ 10 MeV [12, 28, 88].

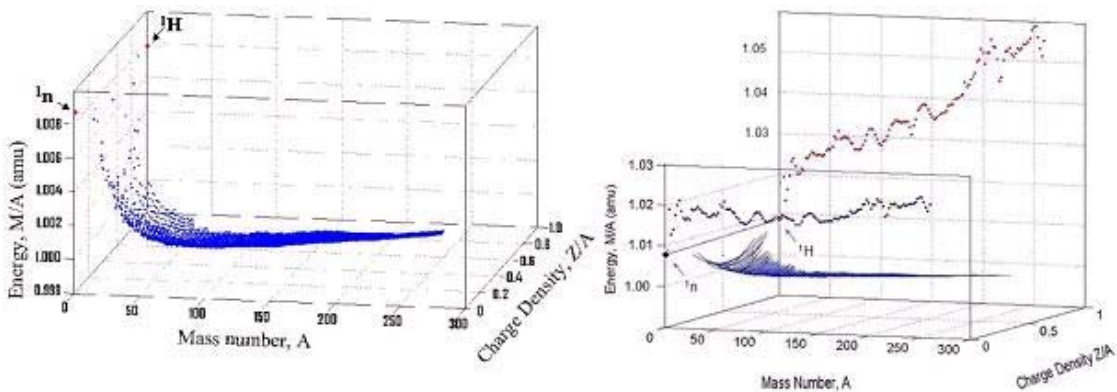

Fig. 8. <u>On the left</u>, the "cradle of the nuclides" [12, 28] shows the potential energy per nucleon for all 2,850 nuclides [85]. <u>On the right</u>, the mass parabolas defined by the data at each value of A are shown, together with their intercepts on the front and back planes at Z/A = 0 and Z/A = 1 [12, 88].

At each value of A>1, the empirically defined mass parabola intercepts the front plane at (Z/A = 0, M/A = $M_{neutron}$ + ~10 MeV) [12, 28, 88]. An example of the mass parabola at A = 27 was published in this journal earlier as Fig. 4 [ref. 28]. Intercept values of M/A at Z/A = 0 for A>150 mu suggest that the potential energy per nucleon in a neutron star may exceed the rest mass of a free neutron by as much as 22 MeV [12].

From these results it was concluded that *repulsive interactions between neutrons are a powerful type of nuclear energy that may be released by neutron emission from neutron stars and other stars that form on them* [12, 28, 88], including the Sun.

Although the prevailing opinion is that neutron stars are "dead" nuclear matter, with neutrons tightly bound at about -93 MeV relative to the free neutron [89], the data shown in Fig. 8 and the intercepts calculated at $Z/A = 0$ suggest that in every case neutrons are "energized" rather than "bound" in assemblages of neutrons at every mass number $A>1$ [12, 28, 88].

The calculated amount of energy released in neutron-emission from a neutron star, ~10-22 MeV per nucleon, exceeds that from fusion or fission reactions. In fission, ~0.1% of the rest mass is released as energy. Fusion of H into He or Fe releases ~0.7% or ~0.8% of the rest mass as energy. Neutron-emission from a neutron star is estimated to release ~1.1% - 2.4% of the neutron's rest mass as energy [12, 28, 88]

These reactions may explain solar luminosity (SL), solar neutrinos, solar mass-fractionation, and the hydrogen-rich solar wind (SW) coming from an iron-rich Sun:

- Neutron emission from the solar core (>57% SL)
    - $<^1n> \rightarrow {}^1n\ +\sim$ 10-22 MeV
- Neutron decay or capture (<5% SL)
    - ${}^1n \rightarrow {}^1H^+ + e^- + $ anti $- \nu\ +$ 0.782 MeV
- Fusion and upward migration of $H^+$ (<38% SL)
    - $4\ {}^1H^+ + 2\ e^- \rightarrow {}^4He^{++} + 2\ \nu\ +$ 27 MeV
- Excess $H^+$ escapes in the solar wind (100% SW)
    - $3 \times 10^{43}\ H^+/yr \rightarrow$ Depart in the solar wind

Most ${}^1H^+$ from neutron-decay is consumed before reaching the solar surface. Only about ~ 1% reaches the surface and is discarded in the solar wind.

CONCLUSIONS

Isotope abundance and mass measurements [2] show that the Sun is an iron-rich plasma diffuser that formed on a collapsed SN core. It consists mostly of elements made near the SN core (Fe, O, Ni, Si, and S), like the rocky planets and ordinary meteorites. Neutron-emission from the central neutron star triggers a series of reactions that generate solar luminosity, solar neutrinos, solar mass-fractionation, and an outpouring of hydrogen in the solar wind. Mass fractionation likely operated in the parent star, and in other stars as well.

These findings lend credence to Birkeland's finding that many solar features resemble those of a magnetized metal sphere [90] and to Richards' suggestion [91] that atomic weights *" . . . tell in a language of their own the story of the birth or evolution of all matter, . . ."* [ref. 91, p. 282]. They also resolve two serious difficulties that Nobel Laureate W. A. Fowler [92] identified in the most basic concepts of nuclear astrophysics:

a.) The solar neutrino puzzle reflects the fact that H-fusion generates <38% of the sun's luminosity [28].

b.) The atomic ratio, O/C ~ 2, at the surface of the Sun because fractionation moves lighter C selectively to the surface. O/C ~ 10 inside the Sun [83].

## ACKNOWLEDGMENTS

We are grateful to the University of Missouri and the Foundation for Chemical Research, Inc. for support and permission to reproduce figures from our earlier reports This paper is dedicated to the memory of Dr. Glenn T. Seaborg**.** who helped organize the 1999 ACS symposium on *The Origin of Elements in the Solar System* [e.g., 37, 38, 48] and sought to resolve some of issues discussed here.